\documentclass{article}
\usepackage{spconf,amsmath,amsfonts,graphicx}
\usepackage{etoolbox,siunitx}
\robustify\bfseries
\usepackage[nolist,nohyperlinks]{acronym}
\usepackage{hyperref}
\usepackage{booktabs}
\newlength{\bibitemsep}
\newlength{\bibparskip}
\let\oldthebibliography\thebibliography
\renewcommand\thebibliography[1]{%
  \oldthebibliography{#1}%
  \setlength{\parskip}{\bibitemsep}%
  \setlength{\itemsep}{\bibparskip}%
}
\usepackage{multirow}

\usepackage{enumitem}
\setlist{nosep, leftmargin=14pt}

\usepackage{mwe} 

\title{PI-RADS v2 Compliant Automated Segmentation of Prostate Zones Using co-training Motivated Multi-task Dual-Path CNN }
\name{Arnab Das* \qquad Suhita Ghosh* \qquad Sebastian Stober}
\address{ Artificial Intelligence Lab (AILab), Otto-von-Guericke-University, Magdeburg, Germany}
\begin{document}
%
\maketitle
 \def\thefootnote{*}\footnotetext{These authors contributed equally to this work}
\begin{abstract}
The detailed images produced by Magnetic Resonance Imaging (MRI) provide life-critical information for the diagnosis and treatment of prostate cancer.
To provide standardized acquisition, interpretation and usage of the complex MRI images, the PI-RADS v2 guideline was proposed.
An automated segmentation following the guideline facilitates consistent and precise lesion detection, staging and treatment.
The guideline recommends a division of the prostate into four zones, PZ (peripheral zone), TZ (transition zone), DPU (distal prostatic urethra) and AFS (anterior fibromuscular stroma).
Not every zone shares a boundary with the others and is present in every slice.
Further, the representations captured by a single model might not suffice for all zones, as observed in \cite{meyer2019towards}.
This motivated us to design a dual-branch convolutional neural network (CNN), where each branch captures the representations of the connected zones separately.
Further, the representations from different branches act complementary to each other at the second stage of training, where they are fine-tuned through an unsupervised loss.
The loss penalises the difference in predictions from the two branches for the same class.
We also incorporate multi-task learning in our framework to further improve the segmentation accuracy.
The proposed approach improves the segmentation accuracy of the baseline (mean absolute symmetric distance) by 7.56\%, 11.00\%, 58.43\% and 19.67\% for PZ, TZ, DPU and AFS zones respectively.   

\end{abstract}
\section{Introduction}
\label{sec:intro}
Prostate cancer (PCa) is the most commonly diagnosed cancer and one of the leading causes of cancer-induced death in men \cite{siegel2021cancer}. 
Regular prostate-specific antigen (PSA) screenings can curb the PCa mortality rate.
However, the screenings do not always provide accurate results and often lead to unnecessary diagnosis and over-treatment \cite{ahmed2017diagnostic}.
\begin{figure}[ht]
\label{fig:seg}
\centering
\includegraphics[scale=0.9]{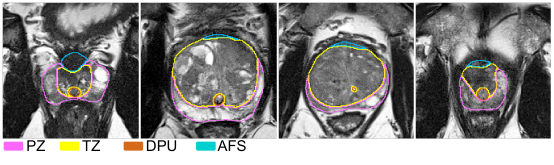}
\caption{Examples of axial slices from prostate T2-weighted MRI, taken from different subjects. Illustrates the variability of the zones across patients.}
\end{figure}
\begin{figure*}[tbph]
  \centering
    \includegraphics[width=\textwidth]{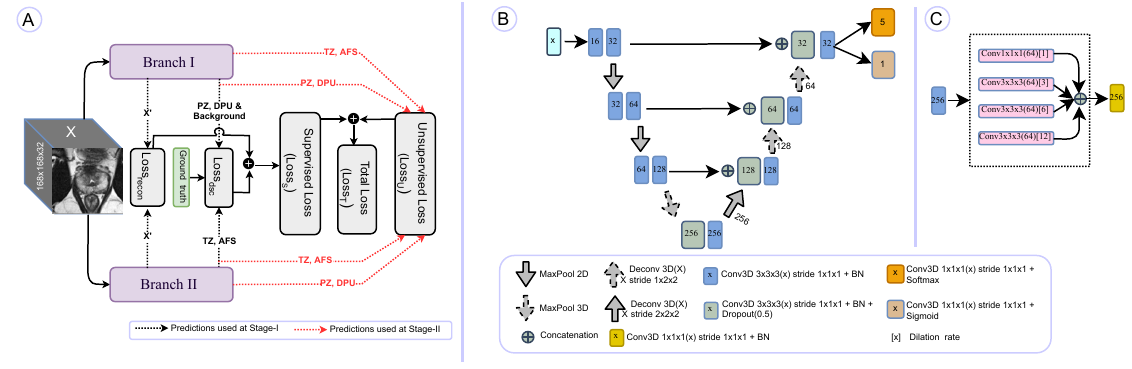}
    \caption{a) An overview of the method proposed. b) The U-Net architecture used in the method. The number inside the blocks represent filter numbers. c) Dilated convolution block used in the mixed model. }
    \label{fig:res}
\end{figure*}
Due to this reason, the high-resolution images produced from multiparametric MRI (mpMRI) are used for clinical assessment, localisation and therapy planning of PCa \cite{turkbey2019prostate}.
To provide guidelines for a standardised acquisition, interpretation and usage of mpMRI, Prostate
Imaging-Reporting and Data System version 2 (PI-RADS v2) \cite{turkbey2019prostate} was introduced.
The guideline considers segmentation of prostate into four anatomical zones, as introduced
by McNeal \cite{vargas2016updated}, shown in Fig.~\ref{fig:seg}.
The segmentation of PZ and TZ facilitates diagnosis and localisation of cancerous cells, as these zones have a higher probability of hosting the clinically significant lesions \cite{moss1994magnetic}.
The delineation of AFS and DPU zones help in the post-diagnostic treatment, dose analysis and focal therapy \cite{meyer2019towards}.
Further, the demarcation of DPU helps in facilitating a precise annihilation of the lesions while sparing the healthy tissue.
However, manual delineation of prostate zones is a time-consuming and error-prone task.
This is due to fuzzy borders, high heterogeneity of pixel intensity within the same zone, and high inter-patient variability, as seen in Fig.~\ref{fig:seg}.
Therefore, an automated segmentation of prostate structures is pertinent to provide a consistent lesion localisation and reduce the cognitive burden on the clinicians.

Many approaches have been proposed for prostate zone segmentation but targeting only PZ and TZ.
Recently, a deep learning-based method \cite{meyer2019towards} was proposed following the PI-RADS v2 recommendation.
The authors proposed a convolutional neural network (CNN) based method using T2-weighted MRI.
The method performed in the range of inter-rater variability for all zones except for AFS, as representations captured by the same model might not be suitable for all zones \cite{meyer2019towards}.
To this end, the representations for AFS are required to be learnt separately.
Further, we can observe in Fig.~\ref{fig:seg} that a pair of zones are directly connected for most of the slices, such as (TZ and AFS) and (PZ and DPU), and some are never connected (AFS and DPU).
Since the connected zones share boundaries, they tend to have similar representations.

In this work, we propose a dual-branch CNN based method, where each branch captures the representations of the connected zones independently, but act complementary to each other.
Further, we perform a two-stage co-training \cite{blum1998combining} motivated training.
In co-training , two views of the data are used to build an initial pair of models, followed by the initially trained models teaching each other.
At the first stage, the branches are trained independently, so that each one captures the representations of the connected zones only.
Subsequently, the representations of each branch is fine-tuned through an unsupervised loss.
The loss is calculated for each zone, which is calculated as the difference between the predictions of the two branches.
We also propose a multi-task loss, which considers the reconstruction of the prostate along with the segmentation of the prostate.
This further facilitates the model to improve the overall segmentation accuracy.
\section{Related work}
\label{sec:relatedwork}

The review \cite{ghose2012survey} summarizes the machine learning and conventional methods for the whole prostate and its (PZ and TZ) zonal segmentation.
The traditional methods are based on deterministic and probabilistic atlas, hybrid methods incorporating intensity and shape prior information.
The review \cite{khan2021recent} provides a detailed overview of the DL methods proposed for prostate zone segmentation.
The PZ and TZ segmentation method proposed in \cite{liu2019automatic} comprised of three sub-networks. 
The authors used a feature-pyramid attention network in the middle to capture minute spatial information in multiple scales from encoded latent images.
\cite{aldoj2020automatic} segmented PZ and TZ zones by an improved U-Net, using the dense-blocks from DenseNet \cite{huang2019convolutional}.
A two stage method was proposed in \cite{singh2020segmentation}, where a probabilistic atlas based approach was applied for PZ and TZ segmentation, followed by the whole prostate segmentation.
Only two methods exist which have targeted four zones.
One of them is a supervised DL method \cite{meyer2019towards}, where an anisotropic 3D U-Net \cite{cciccek20163d} was trained on axial T2-weighted MRI volumes.
They used a combination of isotropic and anisotropic Maxpool layers to cater to the non-isotropic data.
The other one \cite{meyer2021uncertainty} is a semi-supervised method which is a fusion of uncertainty-guided self-training and temporal ensembling.
The method used the annotated data from \cite{meyer2019towards} and a subset of unlabelled data from PROSTATEx challenge dataset \cite{litjens2014computer}.




\section{Methodology}
\label{sec:Method}

In this work, we propose a dual-branch CNN architecture, where the representations of the four zones is learned in two stages of training.
The two-branch concept is based on the hypothesis that it is easier to learn the representations for connected/related zones than all learned together.
Therefore, each branch is trained simultaneously and independent of each other, at the first stage.
This ensures that each branch captures the representations of only connected zones.
At the second stage, the representations from each branch is fine-tuned through an unsupervised loss, which is calculated as the discrepancy between the predictions produced by the two branches, for each zone.
In this way, a transfer of knowledge occurs between the branches, as in co-training.
\cite{caruana1997multitask} showed that multi-task learning (MTL) improves performance in networks, where the network learns to perform multiple tasks simultaneously given a single input for all the tasks. 
This motivates us to incorporate reconstruction loss in the objective to improve the overall segmentation accuracy.
Firstly, we discuss the proposed DL architecture, followed by the two-stage training strategy.
\subsection{DL Architecture}
The prostate zones are extremely dissimilar with respect to shape, texture, inter- and intra-patient variability.
Therefore, the features learned by a single network's filters may not be suitable for segmenting all four zones simultaneously.
However, the connected zones may have similar representations, as they share boundaries.
Therefore, we trained a network with two branches, as shown in Fig~\ref{fig:res}(a).
Branch-I is intended to capture the representations for PZ, DPU and Background, and Branch-II for TZ and AFS.
AFS and TZ are considered in the same branch, as AFS is disconnected from the others except TZ in most of the slices (refer to Fig. \ref{fig:seg}).
Similarly, PZ always contain DPU.
Apart from the zones, there is another class, Background, which contains the pixels outside the prostate.
It was placed in Branch-I containing PZ, as Background share its boundary mostly with PZ.

The AFS zone is the most difficult zone to segment, even for the domain experts \cite{meyer2019towards}.
This is attributed to its extremely indistinct border and widely varying shape and texture across patients.
\cite{yu2015multi} argued that dilated convolution works better for semantic segmentation, due to the increase in the effective receptive field.
Therefore, for the branch having AFS (Branch-II), an additional \textit{dilated} block was added before the first upsampling, as shown in Fig.~\ref{fig:res}(c).
This block contains three dilated convolution layers in parallel with three different dilation rates, which are 3, 6 and 12, along with a 1 x 1 x 1 convolution.
The feature maps are then concatenated and passed through another 1 x 1 x 1 convolution before passing to the upsampling layer in the decoder.
The other branch could also contain the \textit{dilated} block, but it would unnecessarily increase the model parameters.
Although any architecture can be used for the branches, 3D U-Net \cite{cciccek20163d} was considered, shown in Fig. \ref{fig:res}(b).

\subsection{Training Strategy}


The training strategy shown in Fig. \ref{fig:res}(a) can be divided into two stages, Stage-I and Stage-II.
\subsubsection{Stage-I}
At this stage, both the branches are trained in a supervised manner, simultaneously and independently.
To this end, the loss at this stage is computed using the predictions from their relevant branches (PZ, DPU and Background from Branch-I and, TZ and AFS from Branch-II), as shown in Fig. \ref{fig:res}(a).
Eqn.~\ref{eq:sup_loss} shows the loss used at this stage, where $N$ is the total number of voxels, $p_{z,i}$ is the model's prediction and $y_{z,i}$ is the ground truth for the i$^{\textnormal{th}}$ voxel and z$^{\textnormal{th}}$ zone, $Z = \{TZ, PZ, AFS, DPU , Background\}$, and [$\cdot$] is a mask-based indicator function.
Mask M is one when the class predictions $p_{z,i}$ are produced from their relevant branches. 
\vspace{0.5em}
    \begin{equation}
    \label{eq:sup_loss}
    Loss_{dsc} = \sum_{z\in Z} 1  -  \frac{2\sum_{i=1}^N [M_i = 1]p_{z,i} y_{z,i}}{\sum_{i=1}^N [M_i = 1]p_{z,i}^2 + \sum_{i=1}^N [M_i = 1]y_{z,i}^2}
    \vspace{0.5em}
    \end{equation}
The loss function is based on Dice similarity coefficient (DSC), similar to \cite{meyer2019towards}. 
We incorporated multi-task learning (MTL) in the method by using an additional reconstruction loss, as shown in Eqn. \ref{eq:recon_loss}, where $\hat{\textbf{X}}_b$ represents the reconstructed volume by branch $b$, $\textbf{X}$ is the actual prostate MRI volume, and $b \in \{Branch-I, Branch-II\}$.
The loss is based on Structural Similarity Index (SSIM), used typically in reconstruction tasks \cite{2021mrisurvey}.
 \begin{equation}
    \label{eq:recon_loss}  
    Loss_{recon} = \sum_{b}1  -  SSIM(\hat{\textbf{X}}_b , \textbf{X}) 
    \end{equation}
Therefore, the supervised loss ($Loss_{S}$) at this stage is a combination of $ Loss_{recon}$ and $Loss_{dsc}$, as shown in Eqn. \ref{eq:tot_sup_loss}.
 \begin{equation}
   \label{eq:tot_sup_loss} Loss_{S} = Loss_{dsc} + Loss_{recon}
   \end{equation}
\subsubsection{Stage-II}
At this stage, we compute an additional unsupervised loss as shown in Eqn. \ref{eq:unsup_loss}, where $p^{'}_{z,i}$ and $p^{''}_{z,i}$ denote the predictions from Branch-I and Branch-II respectively.
\begin{equation}
      \label{eq:unsup_loss}
      Loss_{U} = \sum_{z \in Z}1  -  \frac{2\sum_{i=1}^N p^{'}_{z,i}p^{''}_{z,i}}{\sum_{i=1}^N p^{'~~2}_{z,i} + \sum_{i=1}^N p^{''~2}_{z,i}} 
       \end{equation}
The loss is computed between the predictions of the branches, which helps in exchange of knowledge, as in co-training.
The loss increases with the disagreement of predictions between the branches.
This acts as a regularizer and helps in reducing the bias induced at Stage-I.
The total loss for this stage is presented in Eqn. \ref{eq:tot_loss}, where the supervised loss restricts the model from catastrophic forgetting \cite{mccloskey1989catastrophic} and the unsupervised loss in generalization.
       
     \begin{equation}
    \label{eq:tot_loss}
    Loss_{T} = Loss_{S} + Loss_{U} 
    \end{equation}

\begin{table*}[htbp]
\footnotesize
\centering
\setlength\tabcolsep{4pt}
 \begin{tabular}[width=\textwidth]{l|cc|cc|cc|cc|cc}
 \toprule
\multirow{2}{*}{Model} & \multicolumn{2}{c}{PZ} & \multicolumn{2}{c}{TZ} & \multicolumn{2}{c}{DPU} & \multicolumn{2}{c}{AFS} & \multicolumn{2}{c}{Zones Avg.} \\

     & {DSC (\%)} & {MAD} & {DSC (\%)} & {MAD} & {DSC (\%)} & {MAD} & {DSC(\%)} & {MAD} & {DSC (\%)} & MAD\\
     \hline
     \hline
        {$M_{base}$}& {75.21} $\pm{0.21}$ & {1.19} $\pm{0.06}$ & {85.87} $\pm{0.42}$ & {1.00} $\pm{0.05}$ & {64.40} $\pm{1.42}$ & {3.44} $\pm{0.46}$ & {39.56} $\pm{1.92}$ & {4.17} $\pm{0.82}$  & {66.26} & {2.45}\\
        \hline
        {$M_{par}$}& {76.43} $\pm{0.59}$ & {1.10} $\pm{0.01}$ & {86.57} $\pm{0.42}$ & {0.95} $\pm{0.04}$ & {64.39} $\pm{3.50}$ & {2.59} $\pm{1.93}$ & {\textbf{42.07}} $\pm{1.46}$ & {3.37} $\pm{0.49}$ & {67.36} & {2.00}\\
        \hline
       {$M_{par\_reco}$}& {\textbf{76.83}} $\pm{0.49}$ & {\textbf{1.07}} $\pm{0.07}$ & {86.93} $\pm{0.26}$ & {0.92} $\pm{0.02}$ & {64.43} $\pm{1.20}$ & {3.30} $\pm{2.42}$ & {40.42} $\pm{1.83}$ & {3.60} $\pm{0.44}$ & {67.15}& {2.22}\\
        \hline
        {$M_{mix}$}& {75.89} $\pm{0.28}$ & {1.14} $\pm{0.03}$ & {86.50} $\pm{0.59}$ & {0.93} $\pm{0.04}$ & {64.20} $\pm{1.95}$ & {2.97} $\pm{2.72}$ & {40.18} $\pm{1.96}$ & {3.91} $\pm{0.29}$ & {66.70}& {2.23}\\
         \hline
        {$M_{mix\_reco}$}& {76.55} $\pm{0.47}$ & {1.10} $\pm{0.06}$ & {\textbf{87.03}} $\pm{0.55}$ & {\textbf{0.89}} $\pm{0.04}$ & {\textbf{65.65}} $\pm{3.09}$ & {\textbf{1.43}} $\pm{2.74}$ & {40.94} $\pm{1.03}$ & {\textbf{3.35}} $\pm{0.34}$ & {\textbf{67.54}}& {\textbf{1.69}}\\
      \hline

\end{tabular}
\caption{Quantitative evaluation for all models. The last column Zones Avg. shows the mean score of all zones.
The best results are in bold.\label{tab:results}}
\end{table*}
\section{Dataset and Experiment Details}
\label{sec:Dataset}

We have used the annotated 98 T2-weighted axial MRI volumes provided by \cite{meyer2019towards}.
To speed up convergence, the voxel intensities were cropped to the first and 99th percentile and then normalized to the range of \([0,1]\).
The train, validation and test split was 58, 20 and 20 respectively.
We have performed 4-fold cross-validation for all our experiments by re-shuffling the volumes from train and validation set.

The supervised state-of-the-art \cite{meyer2019towards} for the prostate zonal segmentation served as the baseline ($M_{base}$).
We denote the proposed two-branch mixed model with MTL as $M_{mix\_reco}$, where the mixed model is the one with different branches, where only one of the branches have dilated blocks.
For ablation study, we trained the following two-branch model variants: with same branches and without MTL ($M_{par}$), with same branches and with MTL ($M_{par\_reco}$), and with different branches without MTL ($M_{mix}$).

The models were trained using ADAM optimizer (learning rate 1e-5).
Each experiment was run with early-stopping (after 30 epochs of no improvement) on the validation set.
The model with the lowest validation loss was selected and used for the evaluation on test data.
To ensure topological correctness, a post-processing step was performed, as done in \cite{meyer2019towards}.
It includes two steps, connected components analysis (CCA) and a signed euclidean distance-based hole filling operation.
The CCA only retains the largest component for each zone, and the latter assigns labels to each label-free voxels, produced by CCA.
Since the zones' predictions come from different branches, a normalization step was performed, before passing the predictions to the post-processing step. 
\section{Results and Discussion}
\label{sec:ResultsandDiscussion}

\begin{figure}[t]

\begin{minipage}[b]{1.00\linewidth}
    \centerline{\includegraphics[scale=0.9]{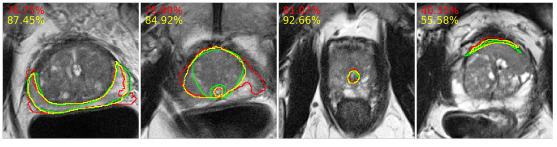}}
\end{minipage}
\caption{Examples of predictions produced for the zones PZ, TZ, DPU and AFS, by $M_{mix\_reco}$ (Yellow) and $M_{base}$ (Red). The ground truth is denoted by Green contour. The mentioned values are DSC scores for the zone prediction from their respective models. }
\label{fig:preds}
\end{figure}

We evaluated the models using DSC and mean absolute symmetric distance (MAD), as done in \cite{meyer2019towards}.
Table \ref{tab:results} portrays the performance of all models.
Considering the overall performance (for all zones), our proposed dual-branch mixed MTL model $M_{mix\_reco}$ outperformed $M_{base}$ with respect to the mean DSC and MAD scores.
A statistical test (one-sided paired t-test with significance level 0.05) showed that $M_{mix\_reco}$ outperformed the baseline for all zones except TZ.
Further, the statistical test resulted in a p-value of 0.0189 (PZ), 0.0517 (TZ), 0.0001 (DPU) and 0.0011 (AFS).
With respect to MAD score also, we obtained similar statistical evidence results.
Interestingly no variant of our proposed two-branch method performed the best for all zones.
But, all the variants of two-branch method outperformed $M_{base}$ for all zones, with respect to both metrics.
Fig. \ref{fig:preds} shows that our proposed model produces segmentation masks closer to the ground truth, compared to $M_{base}$.

For PZ, the two-branch model with MTL ($M_{par\_reco}$) achieved the highest mean DSC score of 76.83\%, which is a 2.15\% increase over $M_{base}$.
Although the mixed model variant $M_{mix\_reco}$ performed closely to $M_{par\_reco}$.
Our proposed model $M_{mix\_reco}$ rectified the over-segmentation of the baseline in many cases, as shown in Fig.~\ref{fig:preds}.
For TZ, $M_{mix\_reco}$ outperformed other variants, where it achieved a mean DSC of 87.03\% which is 1.35\% higher than $M_{base}$.
Similar to PZ, $M_{mix\_reco}$ rectified the over-segmentation of the baseline, as shown in Fig. \ref{fig:preds}.
However, $M_{mix\_reco}$ also over-segmented in many cases.
This is attributed to the similar intensity distribution of the nearby tissue.

\begin{figure}[t]

\begin{minipage}[b]{1.00\linewidth}
    \centerline{\includegraphics[scale=1.0]{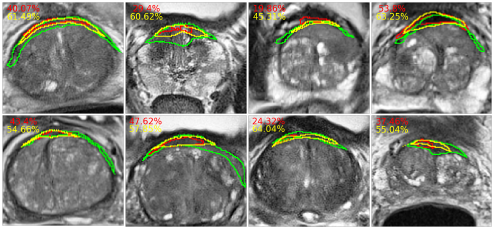}}
\end{minipage}
\caption{More examples of prediction for AFS by $M_{mix\_reco}$ (Yellow) and $M_{base}$ (Red). The ground truth is denoted by Green contour. Images are cropped for visualization. The mentioned values are DSC scores for the zone prediction from their respective models.}
\label{fig:afs}
\end{figure}

Considering the minority classes DPU and AFS, the baseline's mean MAD scores were improved remarkably by 58.43\% (DPU) and 19.67\% (AFS).
This indicated that the proposed approach improved the baseline’s quality of border delineation considerably for the smaller zones.
For DPU $M_{mix\_reco}$ performed the best.
In many cases, both $M_{mix\_reco}$ and $M_{base}$ missed DPU, which is a difficult class to detect due to its severe under-presence in the dataset (less than 1\%).
Interestingly, $M_{par}$ produced the best mean DSC score for AFS (6.34\% better than $M_{base}$).
However, with respect to mean MAD score, $M_{mix\_reco}$ performed the best.
Further, the multi-task model ($M_{par\_reco}$) performed worse than $M_{par}$ for AFS.
This indicates the additional inductive bias introduced by MTL does not always help \cite{caruana1997multitask}, as in the case of AFS.
However, MTL improved the segmentation accuracy for other zones, PZ, TZ, and DPU.
Fig. \ref{fig:afs} shows that our proposed method produced much better segmentation quality for different shapes of AFS, which complies by the distance-based measure's (MAD) value, shown in Tab. \ref{tab:results}.
This indicates the proposed method helps to generalise over variety of shapes observed for AFS.
The code is publicly available on Github.



\section{Conclusion}
\label{sec:Conclusion}

In this work, we presented a co-training motivated dual-branch CNN-based method for simultaneous zonal segmentation of the prostate as per the globally accepted PI-RADS v2 guidelines, from axial T2-weighted MRI volumes.
The method is based on the concept that it is easier to learn representations for similar classes than all considered together.
We also proposed a loss incorporating multi-task learning, which improved the overall segmentation accuracy significantly compared to the baseline method.  
However, the mean DSC score for small regions like AFS is still significantly lower compared to the large regions like TZ and PZ.
One of the reasons being, only 0.3\% of voxels belong to AFS in the dataset, which makes it hard for the model to generalise for such a hard zone with varied shape, size, and appearance.
Therefore, in order to improve the segmentation accuracy of AFS significantly, more good quality annotated data is needed.
Also, smaller structures tend to obtain lesser accuracy for region-based metrics, such as DSC, as mentioned in \cite{meyer2019towards}.
This motivates us to explore other loss functions specifically for the AFS zone,  which are not based on DSC,  as future work.
We did not compare our results to the semi-supervised method \cite{meyer2021uncertainty}, as they used additional 235 unlabelled prostate volumes.
As future work, we will extend our method to include additional unlabeled data.
We also plan to experiment with other perception-aware reconstruction losses used in other imaging modalities \cite{ghosh2021perception}.
\section{Compliance with Ethical Standards}
This research study was conducted retrospectively using human subject data made available in open access by \cite{meyer2019towards}. Ethical approval was not required as confirmed by the license attached with the open access data.

\bibliographystyle{IEEEbib}
\bibliography{strings}

\begin{thebibliography}{10}

\bibitem{meyer2019towards}
Anneke Meyer, Marko Rakr, Daniel Schindele, Simon Blaschke, Martin Schostak,
  Andriy Fedorov, and Christian Hansen,
\newblock ``Towards patient-individual {PI}-rads v2 sector map: {CNN} for
  automatic segmentation of prostatic zones from {T2}-weighted {MRI},''
\newblock in {\em 2019 IEEE 16th International Symposium on Biomedical Imaging
  (ISBI 2019)}. IEEE, 2019, pp. 696--700.

\bibitem{siegel2021cancer}
Rebecca~L Siegel, Kimberly~D Miller, Hannah~E Fuchs, and Ahmedin Jemal,
\newblock ``Cancer statistics, 2021.,''
\newblock {\em CA: a cancer journal for clinicians}, vol. 71, no. 1, pp. 7--33,
  2021.

\bibitem{ahmed2017diagnostic}
Hashim~U Ahmed, Ahmed El-Shater Bosaily, Louise~C Brown, Rhian Gabe, Richard
  Kaplan, Mahesh~K Parmar, Yolanda Collaco-Moraes, Katie Ward, Richard~G
  Hindley, Alex Freeman, et~al.,
\newblock ``Diagnostic accuracy of multi-parametric {MRI} and {TRUS} biopsy in
  prostate cancer ({PROMIS}): a paired validating confirmatory study,''
\newblock {\em The Lancet}, vol. 389, no. 10071, pp. 815--822, 2017.

\bibitem{turkbey2019prostate}
Baris Turkbey, Andrew~B Rosenkrantz, Masoom~A Haider, Anwar~R Padhani, Geert
  Villeirs, Katarzyna~J Macura, Clare~M Tempany, Peter~L Choyke, Francois
  Cornud, Daniel~J Margolis, et~al.,
\newblock ``Prostate imaging reporting and data system version 2.1: 2019 update
  of prostate imaging reporting and data system version 2,''
\newblock {\em European urology}, vol. 76, no. 3, pp. 340--351, 2019.

\bibitem{vargas2016updated}
HA~Vargas, AM~H{\"o}tker, DA~Goldman, CS~Moskowitz, Tatsuo Gondo, Kazuhiro
  Matsumoto, B~Ehdaie, Sungmin Woo, SW~Fine, VE~Reuter, et~al.,
\newblock ``Updated prostate imaging reporting and data system ({PIRADS v2})
  recommendations for the detection of clinically significant prostate cancer
  using multiparametric {MRI}: critical evaluation using whole-mount pathology
  as standard of reference,''
\newblock {\em European radiology}, vol. 26, no. 6, pp. 1606--1612, 2016.

\bibitem{moss1994magnetic}
David~S Moss,
\newblock ``Magnetic resonance imaging of the prostate,''
\newblock in {\em Radiology of the Lower Urinary Tract}, pp. 203--209.
  Springer, 1994.

\bibitem{blum1998combining}
Avrim Blum and Tom Mitchell,
\newblock ``Combining labeled and unlabeled data with co-training,''
\newblock in {\em Proceedings of the eleventh annual conference on
  Computational learning theory}, 1998, pp. 92--100.

\bibitem{ghose2012survey}
Soumya Ghose, Arnau Oliver, Robert Mart{\'\i}, Xavier Llad{\'o}, Joan~C
  Vilanova, Jordi Freixenet, Jhimli Mitra, D{\'e}sir{\'e} Sidib{\'e}, and
  Fabrice Meriaudeau,
\newblock ``A survey of prostate segmentation methodologies in ultrasound,
  magnetic resonance and computed tomography images,''
\newblock {\em Computer methods and programs in biomedicine}, vol. 108, no. 1,
  pp. 262--287, 2012.

\bibitem{khan2021recent}
Zia Khan, Norashikin Yahya, Khaled Alsaih, Mohammed~Isam Al-Hiyali, and Fabrice
  Meriaudeau,
\newblock ``Recent automatic segmentation algorithms of mri prostate regions: A
  review,''
\newblock {\em IEEE Access}, 2021.

\bibitem{liu2019automatic}
Yongkai Liu, Guang Yang, Sohrab~Afshari Mirak, Melina Hosseiny, Afshin
  Azadikhah, Xinran Zhong, Robert~E Reiter, Yeejin Lee, Steven~S Raman, and
  Kyunghyun Sung,
\newblock ``Automatic prostate zonal segmentation using fully convolutional
  network with feature pyramid attention,''
\newblock {\em IEEE Access}, vol. 7, pp. 163626--163632, 2019.

\bibitem{aldoj2020automatic}
Nader Aldoj, Federico Biavati, Florian Michallek, Sebastian Stober, and Marc
  Dewey,
\newblock ``Automatic prostate and prostate zones segmentation of {MRI} using
  densenet-like u-net,''
\newblock {\em Scientific reports}, vol. 10, no. 1, pp. 1--17, 2020.

\bibitem{huang2019convolutional}
Gao Huang, Zhuang Liu, Geoff Pleiss, Laurens Van Der~Maaten, and Kilian
  Weinberger,
\newblock ``Convolutional networks with dense connectivity,''
\newblock {\em IEEE Transactions on Pattern Analysis and Machine Intelligence},
  2019.

\bibitem{singh2020segmentation}
Dharmesh Singh, Virendra Kumar, Chandan~J Das, Anup Singh, and Amit
  Mehndiratta,
\newblock ``Segmentation of prostate zones using probabilistic atlas-based
  method with diffusion-weighted {MR} images,''
\newblock {\em Computer Methods and Programs in Biomedicine}, vol. 196, pp.
  105572, 2020.

\bibitem{cciccek20163d}
{\"O}zg{\"u}n {\c{C}}i{\c{c}}ek, Ahmed Abdulkadir, Soeren~S Lienkamp, Thomas
  Brox, and Olaf Ronneberger,
\newblock ``3d u-net: learning dense volumetric segmentation from sparse
  annotation,''
\newblock in {\em International conference on medical image computing and
  computer-assisted intervention}. Springer, 2016, pp. 424--432.

\bibitem{meyer2021uncertainty}
Anneke Meyer, Suhita Ghosh, Daniel Schindele, Martin Schostak, Sebastian
  Stober, Christian Hansen, and Marko Rak,
\newblock ``Uncertainty-aware temporal self-learning ({UATS}): Semi-supervised
  learning for segmentation of prostate zones and beyond,''
\newblock {\em Artificial Intelligence in Medicine}, vol. 116, pp. 102073,
  2021.

\bibitem{litjens2014computer}
Geert Litjens, Oscar Debats, Jelle Barentsz, Nico Karssemeijer, and Henkjan
  Huisman,
\newblock ``Computer-aided detection of prostate cancer in {MRI},''
\newblock {\em IEEE transactions on medical imaging}, vol. 33, no. 5, pp.
  1083--1092, 2014.

\bibitem{caruana1997multitask}
Rich Caruana,
\newblock ``Multitask learning,''
\newblock {\em Machine learning}, vol. 28, no. 1, pp. 41--75, 1997.

\bibitem{yu2015multi}
Fisher Yu and Vladlen Koltun,
\newblock ``Multi-scale context aggregation by dilated convolutions,''
\newblock {\em arXiv preprint arXiv:1511.07122}, 2015.

\bibitem{2021mrisurvey}
Emmanuel Ahishakiye, Martin~Bastiaan Van~Gijzen, Julius Tumwiine, Ruth Wario,
  and Johnes Obungoloch,
\newblock ``A survey on deep learning in medical image reconstruction,''
\newblock {\em Intelligent Medicine}, 2021.

\bibitem{mccloskey1989catastrophic}
Michael McCloskey and Neal~J Cohen,
\newblock ``Catastrophic interference in connectionist networks: The sequential
  learning problem,''
\newblock in {\em Psychology of learning and motivation}, vol.~24, pp.
  109--165. Elsevier, 1989.

\bibitem{ghosh2021perception}
Suhita Ghosh, Andreas Krug, Georg Rose, and Sebastian Stober,
\newblock ``Perception-aware losses facilitate {CT} denoising and artifact
  removal,''
\newblock {\em 2nd IEEE International Conference on Human-Machine Systems},
  2021.

\end{thebibliography}

\end{document}